\documentstyle[pra,aps,twocolumn,epsf]{revtex}

\begin{document}
\draft
\title{First-order  interference of nonclassical light
\\emitted spontaneously at different times}
\author{Yoon-Ho Kim,\thanks{Email: yokim@umbc.edu} Maria V. Chekhova,\thanks{Permanent address:
Department of Physics, Moscow State University, Moscow, Russia.}
Sergei P. Kulik,$^\dagger$ Yanhua Shih, and Morton H. Rubin}
\address{Department of Physics, University of Maryland, Baltimore
County, Baltimore, Maryland 21250}
\date{Submitted to Phys. Rev. A}

\maketitle

\begin{abstract}
We study first-order interference in spontaneous parametric
down-conversion generated by two  pump pulses that do not overlap
in time.  The observed modulation in the angular distribution of
the signal detector counting rate can only be explained in terms
of a quantum mechanical description based on biphoton states. The
condition for observing interference in the signal channel is
shown to depend on the parameters of the idler radiation.
\end{abstract}

\pacs{PACS Number: 42.50.Dv, 03.65.Bz}

Nonclassical interference is one of the most remarkable phenomena
of quantum optics. In particular, it can be observed in
experiments with spontaneous parametric down conversion (SPDC)
\cite{Photons}, a nonlinear optical process in which
higher-energy pump photons are converted into pairs of
lower-energy photons (usually called signal and idler) inside a
crystal with quadratic nonlinearity.  It has been shown that the
state of the signal-idler photon pair is entangled in space-time,
polarization, or both \cite{entanglement}. Due to the
nonclassical correlation between the signal and the idler photons
emitted in SPDC, the term ``biphoton" has been suggested
\cite{biphoton}.  Many experiments made use of SPDC to
demonstrate fascinating topics in quantum optics, such as Bell's
inequalities violation, quantum communication, quantum
teleportation, etc \cite{spdc}. All these experiments  belong to
basically the same category: quantum interference. The existence
of the SPDC interference enables one to monitor the structure of
biphoton fields. This effect can be used in quantum
communication, computation, and cryptography \cite{comp}.

Among the variety of interference experiments, there is a large
group of works where first-order interference is observed in
signal or idler SPDC radiation emitted from spatially separated
domains \cite{Mandel,Herzog,Burlakov}. This kind of interference
is nonclassical. Indeed, each of the signal and idler beams has
noise (thermal) statistics; from a classical viewpoint, spatially
separated SPDC sources should exhibit no interference pattern in
the signal or idler beams. One cannot observe stable first-order
interference of light emitted by independent classical thermal
sources such as, for instance, two similar light-emitting
diodes\cite{diodes}. To explain the interference observed for
SPDC radiation, one should take into account that each signal or
idler photon is generated not from some particular point but from
the whole volume with quadratic nonlinearity, pumped by a
coherent pump. And indeed, in this case one cannot tell whether a
photon was born at one point or another - the interference is
observed in agreement with Feynman's indistinguishability
criterion\cite{Feynman}. Another interesting feature of the
first-order interference of SPDC is that it depends on the phases
of  pump, signal, and idler waves, and one can speak of
``three-frequency'' interference \cite{Burlakov2}. Moreover, the
condition for the interference to be observed in, say, signal
radiation depends on the parameters of the idler
radiation\cite{Burlakov}.

In this Letter, we report an experimental observation of a novel
type of quantum interference, in which the sources of two-photon
radiation are separated not in space but in time, as SPDC is
generated from a train of coherent femtosecond
pulses~\cite{Keller}. Depending on the experimental conditions,
interference can be observed either in the angular distribution
of the signal intensity (first-order interference measured by
single-detector method) or in the coincidences of photocounts
from the detectors registering the signal and the idler radiation
(second-order interference measured by two-detector method). In
this paper, we focus on the first-order interference;
second-order interference effects will be discussed elsewhere.

Let us consider type-II SPDC field~\cite{type} generated in a crystal
of length $L$ from a train of two short pump pulses. The signal
radiation is separated from the idler one (Fig.~1) and its intensity
is measured by a detector that selects a sufficiently narrow
frequency band and a sufficiently small solid angle. (The meaning of
the words `sufficiently small' will be clear from further
consideration.)

The pump field dependence on time can be represented in the form

\begin{equation} E_p(r,t)=
\widetilde{E}(t-z/u_p)\exp (-i\omega_pt+ik_pz), \label{2.3}
\end{equation}
where $u_p$ is the pump group velocity, $\widetilde{E}(t)$ is the
envelope, and $\omega_p$ is the pump central frequency. It is
supposed that the spectral band of the pump is much less than
$\omega_p$ (quasi-monochromatic case). If the pump consists of two
identical pulses separated in time by $T_p$, then
$\widetilde{E}(t)=E_0(t)+E_0(t+T_p)\exp(-i\omega_pT_p)$, where
$E_0(t)$ is a single-pulse envelope.

In the first order perturbation theory, the quantum state of SPDC
is given by \cite{Photons}

\begin{equation}
|\Psi\rangle= |vac\rangle+\sum_{k_s, k_i}F_{k_s,
k_i}|1\rangle_{k_s}|1\rangle_{k_i}, \label{2.4}
\end{equation}
where $F_{k_s, k_i}$ is the two-photon spectral function,
\begin{eqnarray}
F_{k_s, k_i}&=&-
\frac{i\chi}{\hbar}\int_{t_0}^tdt'\int_Vd^3r\widetilde{E}(t'-z/u_p)\nonumber
\\&&\times \exp \{-i(\omega_p- \omega_s-\omega_i)t'+i({\bf k}_p-{\bf
k}_s-{\bf k}_i)\cdot{\bf r}\}, \label{2.5}
\end{eqnarray}
and the notation $|1\rangle_{k_s}|1\rangle_{k_i}$ means a two-photon
state in the modes ${\bf k}_s, {\bf k}_i$.

Since the pump pulse is bounded in time, the integration over
$t'$ can be extended to infinite limits. As a result, it gives
the Fourier transform of the pump envelope,
$\widetilde{E}(\omega_s+\omega_i-\omega_p)$, which in the cw case
becomes a $\delta$-function,
$\delta(\omega_s+\omega_i-\omega_p)$. We obtain

\begin{eqnarray}
F_{k_s,k_i}&=& -\frac{2\pi i\chi}{\hbar}
\widetilde{E}(\omega_s+\omega_i-\omega_p)\delta(k_{sx}+k_{ix})
\nonumber
\\ &&\times
\hbox{sinc} \left\{\frac{L}{2}\left(k_p-k_{sz}-k_{iz}+
\frac{\omega_s+\omega_i-\omega_p}{u_p}\right)\right\},
\label{2.7}
\end{eqnarray}
where, for example, $k_{s,x}$ and $k_{s,z}$ are the transverse
and longitudinal components of the signal wavevector,
respectively. For a two-pulse pump, the envelope spectrum is
$\widetilde{E}(\omega)=E_0(\omega)\hbox{cos}\left\{(\omega-\omega_p)\frac{T_p}{2}\right\}$,
with $E_0(\omega)$ denoting the Fourier transform of the
single-pulse envelope.

The probability of detecting a biphoton is $P_c= |F_{k_{s},
k_{i}}|^2$. Since we are dealing with a two-photon state, the
probability $P_s$ of a {\it photocount from the signal detector} is
calculated by integrating $P_c$ over all idler modes. Thus, the
photon counting rate in the signal detector is

\begin{eqnarray}
R_s&\sim &\int dk_{iz} dk_{ix}|F_{k_s, k_i}|^2 \nonumber \\&=
&\frac{4\pi^2\chi^2 L^2}{\hbar^2} \int dk_{iz}
\left|E_0\left(\omega_s+ \omega_i(k_i)- \omega_p\right)\right|^2
\nonumber \\
&\times&\hbox{cos}^2\left\{(\omega_s+\omega_i(k_i))\frac{T_p}{2}\right\}
\nonumber\\&\times&
\hbox{sinc}^2\left\{\frac{L}{2}\left(k_p-k_{sz}-k_{iz}+
\frac{\omega_s+\omega_i(k_i)-\omega_p}{u_p} \right)\right\},
\label{2.8}
\end{eqnarray}
where $\omega_i(k_i)$ is the dispersion dependence for the idler
beam and $k_{ix}=-k_s\hbox{cos}\theta_s$, with $\theta_s$
denoting the signal angle of scattering.

The cosine modulation in Eq.~(\ref{2.8}) will not be averaged out
by the integration over $k_{iz}$ if the squared sinc function is
much narrower than this modulation. In this case the squared sinc
acts as a delta-function in the integral, thus

\begin{eqnarray}
R_s(\omega_s,\theta_s)&\sim&
\left|E_0\left(\omega_s+\omega_i(k_{i})- \omega_p\right)\right|^2
\nonumber \\
&&\times\hbox{cos}^2\left\{(\omega_s+\omega_i(k_{i}))\frac{T_p}{2}\right\}.
\label{2.9}
\end{eqnarray}

Clearly, Eq.(\ref{2.9}) gives a modulated  structure for the
counting rate of the signal detector. This modulation can be
observed in several ways. One can observe interference by varying
$\omega_s$ or by varying $T_p$, which  both enter the squared
cosine in Eq.(\ref{2.9}). In this work, however, we obtain the
modulation by varying the signal angle of scattering. This is
possible since $k_{i}$, hence $\omega_i(k_{i})$, actually depends
on $\theta_s$. This can be shown by expanding $\omega_i$ in the
vicinity of $\omega_p/2$.

The sinc-square function in the integral (\ref{2.8}) is much narrower
than the cosine modulation if

\begin{equation}
\frac{1}{d\omega_i/dk_{zi}}\gg\frac{T_p}{L}\frac{1}
{(-1+\frac{1}{u_p}d\omega_i/dk_{zi})}.
\end{equation}

\noindent For near-collinear scattering, we can take
approximately $d\omega_i/dk_{zi}=d\omega_i/dk_i\equiv u_i$.
Hence, we obtain the following condition for observing the
first-order interference in SPDC from a two-pulse pump:

\begin{equation}
Q\equiv\frac{L(u^{-1}_p-u^{-1}_i)}{T_p} \gg 1.
\label{2.13}
\end{equation}

The other condition is the assumption we have used when obtaining
Eq.~(\ref{2.9}): the signal detector should select a sufficiently
narrow frequency band and a sufficiently small solid angle. Indeed,
 the interference structure will be wiped out
if Eq.~(\ref{2.9}) is integrated over a broad band of signal
frequencies or angles of scattering. Thus, the requirement to the
frequency band of the signal detector is

\begin{equation}
\Delta\omega_s\ll\frac{\pi}{T_p}. \label{2.14}
\end{equation}

In the experiment, the pump is the frequency-doubled radiation
from a mode-locked Ti:Sapphire laser with initial central
wavelength $800$nm. After frequency doubling, the pulse duration
is $140$fsec, and the repetition rate is $90$MHz. The pump pulse
is then fed into the polarization pulse splitter consisting of
two Glan prisms G and a set of quartz rods QR placed between the
prisms (Fig.~1). The axes of the Glan prisms are parallel to the
pump polarization. The `fast' and `slow' axes of the quartz rods
lie in the plane normal to the pump beam and are directed at
$45^\circ$ to the pump polarization. Due to the birefringence of
the quartz rods, at the output of the polarization splitter each
pump pulse is transformed into two pulses that have equal
amplitudes but are delayed in time with respect to one another by
$L_q(u_o^{-1}-u_e^{-1})$, where $L_q$ is the total length of the
quartz rods and $u_o$, $u_e$ are group velocities of the ordinary
and extraordinary waves in quartz at the pump wavelength
($400$nm). The SPDC radiation is generated in a BBO crystal cut
for collinear frequency-degenerate type-II phase matching. The
SPDC radiation is separated from the pump radiation by means of a
prism (P) and a pinhole. A polarizing beam splitter (PBS)
separates the signal radiation from the idler one, and the idler
beam is discarded. The detector (an avalanche diode operating in
the Geiger mode) is placed at the focal plane of a lens (F =
$20$cm), so that the transverse displacement of the detector is
proportional to the angle of scattering, $x\sim F\theta_s$.  In
front of the detector, we place one of the three narrow-band
filters (IF) with central wavelength $\lambda_s = 2\lambda_p=
800$nm and bandwidths $\Delta\lambda_s =  1, 3$, and $10$nm,
respectively, for different measurements. The intensity of the
signal radiation is measured as a function of the angle of
scattering, $I_s(\theta_s)$. The angle is scanned by using a step
motor (SM) that moves the detector in the focal plane of the
lens. The parameters are bandwidth of the filter,
$\Delta\lambda_s$, and the time delay between the two pump
pulses, $T_p$, which is varied by using quartz rods with total
length $L_q = 20$, $12.5$, and $7.5$mm, corresponding to the
delays $T_p = 744$, $465$, and $279$fsec, respectively.

The fact that each pair of pulses is actually repeated at a rate
of 90 MHz leads to a fine structure in the single-counting
distribution (\ref{2.9}), which is much narrower than the
bandwidth of the filters we use in our experiment. Therefore, it
is not observable.

To test Eq.~(\ref{2.13}), we use a BBO crystal of length $3$mm to
generate SPDC. In Fig.~2, the intensity is plotted versus the
detector displacement for all three delays $T_p$. All
dependencies are obtained with the $1$ nm interference filter.
Note that in this case, condition (\ref{2.14}) is satisfied for
all three delays: $\pi/\Delta\omega_s\sim 2000$~fs. However, the
interference visibility in all plots is different. The highest
visibility of the interference pattern is observed for the
smallest time interval between the pulses, $279$fsec, with $Q\sim
3$ [Fig.~2(a)]. For the intermediate delay, $465$fsec, the
interference pattern is observed with lower visibility
[Fig.2(b)]. In this case, $Q\sim 2$. Evidently, Eq.~(\ref{2.13})
is not satisfied for the largest delay $744$fsec ($Q\sim 1$),
therefore, in this case the interference structure completely
vanishes [Fig.~2(c)]. For comparison, the angular spectrum of
SPDC in the case of a single-pulse pump is shown in Fig.2(d). In
agreement with Eq.~(\ref{2.9}), the modulation period is larger
for smaller delays. In all experimental plots, positions of the
oscillation peaks are determined by the delay introduced between
the pump pulses, in perfect agreement with the theoretical
calculation [shown by arrows in Figs.~2(a),(b),(c)]. However, the
theory discussed above does not give explicit description of the
observed asymmetry of the angular spectral envelope. In this
Letter, we only focus on the interference modulation; the shape
of the angular spectrum envelope will be considered elsewhere.

The spectral width of the filter, $\Delta\lambda_s$, has a strong
influence on the interference pattern. Changing the filter bandwidth
from $1$nm to $3$nm, we observed a considerable decrease of the
visibility. At $\Delta\lambda_s=10$nm, no interference structure was
observed, in agreement with condition~(\ref{2.14}).

Let us give a physical interpretation of condition (\ref{2.13}) for
observing the first-order interference in SPDC signal radiation from
a two-pulse pump. Note that although it is the signal radiation that
one detects, Eq.~(\ref{2.13}) contains only the pump and the idler
parameters.

Considering only the signal radiation, it would seem that the
only condition for the first-order interference to take place is
Eq.~(\ref{2.14}), which states that the filter inserted in front
of the signal detector should have smaller bandwidth compared to
the pump spectrum modulation. Indeed, if the signal photon
wavepackets are spread in time by more than $T_p$, the signal
photons born from different pump pulses are at first sight
indistinguishable. However, it is worth remembering that the
indistinguishability criterion should be understood as
indistinguishability {\it in principle. In principle}, we could
equip our setup by an idler detector with a broad-band filter and
register photocounts from the idler detector (Fig.~3). Then for
each signal photon, detection of its twin idler photon is well
localized in time with respect to the pump pulses, which could
mean that we can always distinguish between a pair born from the
first pulse and a pair born from the second pulse. Let us recall
now that the BBO crystal has finite length $L$. Then a photocount
in the idler detector can appear delayed from the corresponding
pump pulse by any time $0<t<\Delta t_i, \Delta t_i=
L(u_i^{-1}-u_p^{-1})$ (Fig.~3). Idler photocounts from different
pump pulses become indistinguishable if $\Delta t_i\gg T_p$, and
we obtain the second necessary condition for the interference,
which is Eq.~(\ref{2.13}).

There is an analogy between the first-order interference observed for
SPDC generated from two spatially separated domains and for SPDC
generated from two separate pump pulses. Indeed,
condition~(\ref{2.13}) ensures that the crystal is long enough so
that an idler photon generated by the first pump pulse can meet the
second pulse (see Feynman diagram in Fig.~3). In the case of
spatially separated SPDC sources, first-order
interference~\cite{Mandel,Burlakov} is possible when idler waves
propagate through both spatial domains where SPDC take
place~\cite{noinduced}. Similarly to Ref.~\cite{Burlakov}, where the
effect has simple explanation in terms of the pump angular spectrum,
here it can be explained by the cosine modulation of the pump
frequency spectrum. Condition~(\ref{2.13}) has the following spectral
interpretation: the typical scale of the pump spectrum modulation
should be much larger than the width of the idler radiation spectrum,
which is determined by the length of the crystal~\cite{Photons}.

In conclusion, we have demonstrated the first-order interference of
nonclassical light generated from two pump pulses well separated in
time. The interference is explained by a quantum mechanical
calculation in terms of biphoton states. The interference pattern is
observed in the angular distribution of the signal intensity.
Interference takes place if the following condition is satisfied: the
time indeterminacy of the delay between the idler photon and the
corresponding pump pulse is much larger than the time interval
between the pump pulses. From the spectral viewpoint, this condition
means that the modulation of the pump spectrum, determined by the
distance between the pulses, should be much larger than the width of
the idler radiation spectrum from a cw pump, determined by the
crystal length. Thus, the interference visibility is sensitive to the
crystal length. It is also sensitive to the spectral width of the
narrow-band filter used for the frequency selection of the signal
radiation.

This work was supported in part by The Office of Naval Research
and an ARO-NSA grant. MVC and SPK also acknowledge partial
support from the Russian Foundation for Basic Research, grant No.
97-02-17498.

\begin{figure}[tbp]
\centerline{\epsfxsize=3in \epsffile{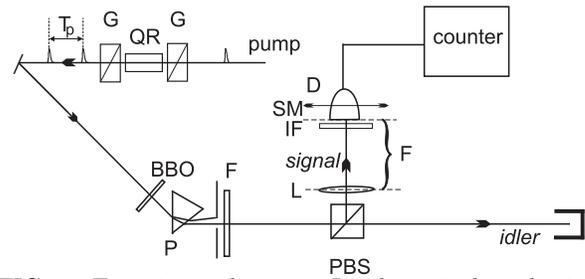}}
\caption{Experimental setup. Incident single pulse is divided
into two temporally separated pulses ($T_p$) by the use of two
Glan prisms and a set of quartz rods. Collinear degenerate
type-II SPDC is generated from the BBO crystal pumped by the two
pulses. The angle of scattering is scanned by moving a detector
(D) in the focal plane of the lens. Interference filters (IF)
with bandwidth 1nm, 3nm, and 10nm are used for spectral selection
of the signal.}\label{fig:figure1}
\end{figure}

\begin{figure}[tbp]
\centerline{\epsfxsize=3in \epsffile{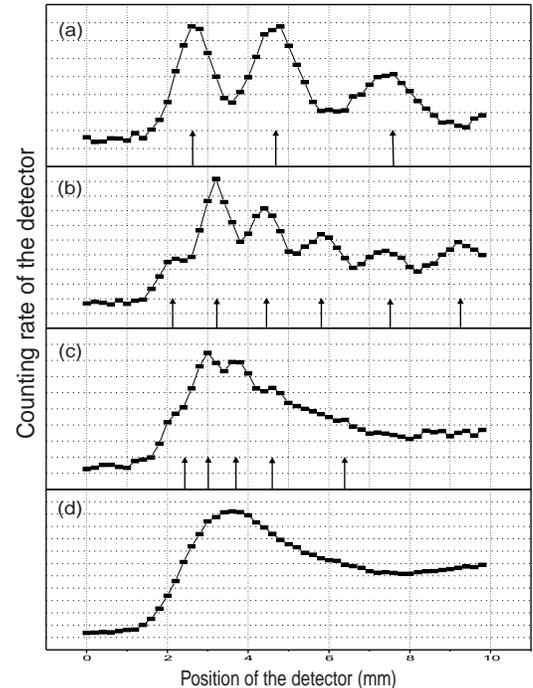}} \caption{The
angular spectrum  of SPDC intensity for three values of $T_p$:
(a) $279$fsec, (b) $465$fsec, (c) $744$fsec. The angle of
scattering is connected with the position of the detector $x$ as
$\theta_s=\frac{x}{F}$, $F=20$cm. The bandwidth of the
interference filter is $1$nm for all three plots. Arrows indicate
calculated positions of interference maxima. The plot (d) shows
the angular spectrum of SPDC in the case of a single pump
pulse.}\label{fig:figure2}
\end{figure}

\begin{figure}[tbp]
\centerline{\epsfxsize=3in \epsffile{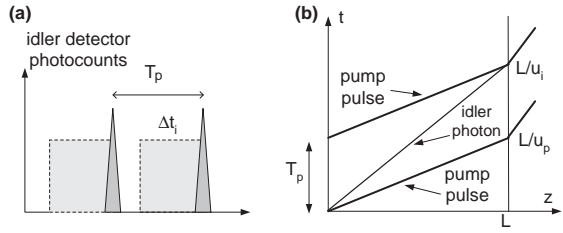}} \caption{(a)
photocounts from a broad-band idler detector can be used to
distinguish between signal photons generated from different pump
pulses. Each idler photocount comes after the corresponding pump
pulse, with the time indeterminacy of $\Delta
t_i=L(u_i^{-1}-u_p^{-1})$ (shown by a grey square). If $\Delta
t_i < T_p$, signal photocounts from different pump pulses are
distinguishable. (b) Feynman diagram illustrating the delay
between the pump pulse and the idler photon inside the BBO
crystal.} \label{fig:figure3}
\end{figure}


\begin{references}

\bibitem{Photons} D.N. Klyshko, {\em Photons and Nonlinear Optics}
(Gordon \& Breach, New York, 1988).

\bibitem{entanglement} Y.H. Shih, A.V. Sergienko, M.H. Rubin, T.E. Kiess,
 and C.O. Alley, Phys. Rev. A {\bf 50}, 23 (1994);
 Y.H. Shih and A.V. Sergienko, Phys. Rev. A {\bf 50}, 2564
(1994); M.H. Rubin, D.N. Klyshko, Y.H. Shih, and A.V. Sergienko,
Phys. Rev. A {\bf 50}, 5122 (1994); P.G. Kwiat, K. Mattle, H.
Weinfurter, A. Zeilinger, A.V. Sergienko, and Y.H. Shih, Phys.
Rev. Lett. {\bf 75}, 4337 (1995).

\bibitem{biphoton} D.N. Klyshko, Soviet Physics-JETP {\bf 56}, 755 (1982).
\bibitem{spdc} Y.H. Shih and C.O. Alley, Phys. Rev. Lett. {\bf 61},
2921 (1988); K. Mattle, H. Weinfurter, P.G. Kwiat, and A.
Zeilinger , Phys. Rev. Lett. {\bf 76}, 4656 (1996); J. Brendel,
N. Gisin, W. Tittel, and H. Zbinden, Phys. Rev. Lett. {\bf 82},
2594 (1999); D. Bouwmeester, J.-W. Pan, K. Mattle, M. Eidl, H.
Weinfurter, and A. Zeilinger, Nature {\bf 390}, 575 (1997); D.
Boschi, S. Branca, F. De Martini, L. Hardy, and S. Popescu, Phys.
Rev. Lett. {\bf 80}, 1121 (1998).

\bibitem{comp} C.H. Bennett and S.J. Wiesner, Phys. Rev. Lett. {\bf 69},
2881 (1992); A.K. Ekert, J.G. Rarity, P.R. Tapster, and G.M.
Plama, {\it ibid.} {\bf 69}, 1293 (1992); J.D. Franson and H.
Ilves, J. Mod. Opt. {\bf 41}, 2391 (1994).

\bibitem{Mandel} X.Y. Zou, L.J. Wang, and L. Mandel, Phys. Rev. Lett.,
{\bf 67}, 318 (1991).

\bibitem{Herzog} T.J. Herzog, J.G. Rarity, H. Weinfurter, and
A. Zeilinger, Phys. Rev. Lett. {\bf 72}, 629 (1994).

\bibitem{Burlakov} A.V. Burlakov, M.V. Chekhova, D.N. Klyshko,
S.P. Kulik, A.N. Penin, Y.H. Shih, and D.V. Strekalov, Phys. Rev. A {\bf 56}, 3214
(1997).

\bibitem{diodes} In this case, even if a narrow-band filter is
used, the interference pattern is stable only for the time
$2\pi/\Delta\omega$, where $\Delta\omega$ is the frequency bandwidth
of the filter

\bibitem{Feynman} R. Feynman, R. Leighton, and M. Sands, {\it The
Feynman Lectures on Physics}, Vol.III, Addison Wesley, Reading
(1965).

\bibitem{Burlakov2} A.V. Burlakov, S.P. Kulik, A.N. Penin, and M.V.
Chekhova, JETP {\bf 86}, 1090 (1998).

\bibitem{Keller}Second-order interference effects from a
pump consisting of several short pulses were described in T.E.
Keller, M.H. Rubin, and Y.H. Shih, Phys. Lett. A {\bf 244}, 507
(1998) and observed in Y.-H. Kim, M.V. Chekhova, S.P. Kulik, and
Y.H. Shih, Phys. Rev. A {\bf 60}, R37 (1999).

\bibitem{type} In type-I
SPDC, the signal and idler photons have the same polarization, in
type-II SPDC they have orthogonal polarization.

\bibitem{noinduced} Of course we do not mean that the idler
photons created from the first pulse somehow influence the SPDC
process from the second pulse, like in the spatial case the idler
radiation emitted from the first domain does not influence the
process in the second domain. The output photon numbers in both
signal and idler modes are so small that all induced effects can be
neglected.

\end{references}
\end{document}